\documentclass[preprint2]{aastex}
\shortauthors{SINGAL}
\received{}
\begin{document}
\title{RELATIVISTIC DOPPLER BEAMING AND MISALIGNMENTS IN AGN JETS}
\author{ASHOK K. SINGAL}
\affil{ASTRONOMY \& ASTROPHYSICS DIVISION, PHYSICAL RESEARCH LABORATORY, \\
NAVRANGPURA, AHMEDABAD - 380 009, INDIA; asingal@prl.res.in}
\begin{abstract}
Radio maps of AGNs often show linear features, called jets, both on pc as well as kpc scales. These jets supposedly  
possess relativistic motion and are oriented close to the line of sight of the observer and accordingly the relativistic Doppler 
beaming makes them look much brighter than they really are in their respective rest-frames. The flux boosting due to the relativistic 
beaming is a very sensitive factor of the jet orientation angle, as seen by the observer. 
Quite often large bends are seen in these jets, with misalignments being $90^\circ$ or more and might imply a change in the orientation angle  
that could cause a large change in the relativistic beaming factor. Such large bends 
should show high contrasts in the brightness of the jets, before and after the misalignments, if relativistic 
beaming does play an important role in these jets. It needs to be kept in mind that sometimes a small intrinsic change in the jet angle might appear 
as a much larger misalignment due to the geometrical projection effects, especially when seen close to the line of sight. Of course what really 
matters is the final orientation angle of the jet with respect to the observer's line of sight. Taking the geometrical projection effects 
properly into account, we calculate the consequences of the presumed relativistic beaming and demonstrate that there ought to be large brightness 
ratios in jets before and after the observed misalignments.
\end{abstract}
\section{\bf INTRODUCTION}
Radio galaxies and quasars, belonging to the genus Active Galactic Nuclei (AGNs), often show linear features called jets, which presumably are 
the channels 
of relativistic plasma through which energy is  continually transported to outer parts of these AGNs. There is evidence enough that these jets are 
relativistic, at least in quasars and radio galaxies of type FR II (Fanaroff \&  Riley 1974) and relativistic Doppler beaming could be an important 
factor in their appearance to the observer. The Lorentz factors could be high, $\gamma\sim 5 - 40$, as estimated from the observed superluminal motion 
(Cohen et al. 1977; Kellermann et al. 2003; Jorstad et al. 2005; Marscher 2006). There is other, independent, evidence for the relativistic Doppler 
beaming, from the high brightness temperatures ($T_b$) inferred from the short period variability. 
The estimated $T_b$ values exceed the theoretical limit of $\sim 10^{12}$, initially thought to be set by the large inverse Compton 
losses at still higher $T_b$ (Kellermann \& Pauliny-Toth 1969), therefore called in literature for long as an inverse Compton limit, 
though of late a somewhat stricter limit $\sim 10^{11.3}$ has instead been shown 
to be set by the diamagnetic effects in a synchrotron source which lead to the condition of equipartition among radiating charges and 
the magnetic field and which is also the configuration of  minimum energy for the source (Singal 1986; 2009). 
But much larger brightness temperatures, violating the above incoherent brightness temperature limit, have been inferred 
for the centimeter variable sources. This
excess in brightness temperatures has been explained in terms of a bulk relativistic motion of the emitting component (Rees 1966; 
Blandford \& K\"onigl 1979). The relativistic Doppler factors required to explain the excessively high  
temperatures up to $\sim 10^{19}$~K (Quirrenbach et al. 1992; Wagner \& Witzel 1995) for the intra-day variables 
are $\delta \stackrel{>}{_{\sim}} 10^2$. Thus the evidence for relativistic flows and relativistic beaming in AGNs is quite strong. 
The flux boosting due to relativistic beaming is a very sensitive factor of the orientation angle $\theta$ of 
the jet with respect to the line of sight to the observer. A slight change in $\theta$ could cause a very large 
change in the observed flux density. The one-sidedness of jets seen in many AGNs is explained by the  
difference in the relativistic beaming on the two sides because of their different orientations with respect to the observer's line of sight. 

Now what appears mysterious is that quite often large bends are seen in these jets, with misalignments being $90^\circ$ or even more 
(Pearson \& Readhead 1988; Conway \& Murphy 1993; Appl et al. 1996; Kharb et al. 2010) and might imply a change 
in the orientation angle, which would cause a large change in the relativistic beaming factor. At least in many cases these large bends 
are not accompanied by high contrasts in the brightness of the jets before and after the bends. Some examples are: 3C309.1 (Wilkinson et al. 1986), 
1823+568 (O'dea et al. 1988), 3c66A, 0528+134, 1803+784, BL LAC (Jorstad et al. 2005) and S5 0716+714 (Rani et al. 2015).  
These all may not be consistent with the relativistic beaming models. However, no systematic statistical study has been done about the brightness changes 
in the jet after a misalignment to make an unambiguous statement. In fact there is no statistically unbiased study available about the absolute frequency 
of occurrence of bending in a  complete sample. 

It needs, however, to be kept in mind that sometimes a small bending angle might appear 
as a much larger misalignment due to the geometrical projection effects, especially when seen close to the line of sight. 
The argument goes like this. 
Let $\theta$ be the angle that the jet initially makes with the line of sight and let $\eta$ be the misalignment angle as seen by the observer 
in the sky plane (perpendicular to the line of sight).  
Then the misalignment must have a component perpendicular to the initial direction of the jet (if not then no misalignment 
would be noticed in the jet). If $\zeta$ is the change in angle at the source, then due to foreshortening of the parallel component by $\sin\theta$ 
when projected in the sky plane, we get
\begin{equation}
\label{eq:p59.0}
\tan \eta = \tan \zeta/ \sin \theta \sim  \gamma\tan \zeta,
\end{equation}
for $\sin \theta \sim 1/\gamma$ (assuming a relativistic beaming with $\gamma$ as the Lorentz factor).  
Thus the misalignment of the jets will appear enhanced by a factor $1/ \sin \theta \sim \gamma$ in relativistic beaming cases.
As an example, a $3^\circ$ bend could appear as a $30^\circ$ misalignment for a $\gamma=10$ case. 
However, as much larger misalignments ($\eta \stackrel{>}{_{\sim}} 90^\circ$) have been seen, then one would still need reasonably large $\zeta$ 
in order to explain the observations (unless $\theta \sim 0$). Of course what really 
matters is the final orientation angle $\theta_1$ of the jet with respect to the observer's line of sight. For that, one has to evaluate the 
projection effects in a more precise and rigorous manner and we shall endeavor to do so here. Accordingly, we shall  
explore the question what relativistic beaming models predict about the expected contrast in the 
jet brightness before and after the observed misalignments, taking into account proper geometrical projection effects. 
\section{GEOMETRY OF THE JET BENDING}
Following Conway \&  Murphy (1993), we assume a simple jet bending model where there is a change only 
in the direction of the jet motion (for simplicity we take the bending to be a sudden discontinuous change and not a gradual turning of the jet).
Speed of the jet material is assumed to remain constant during the bending and we further assume that there is no change in 
the intrinsic properties (in particular the intrinsic intensity of the jet material) before and after the bending.
This is for the number of free variables required to explain the observations to be kept at a minimum. 
A change in the jet speed, with a corresponding change in the relativistic Lorentz factor, alone  
would not cause any change in the apparent direction of the jet seen by the observer, though the brightness could change substantially depending 
upon the change in the jet speed. A change in the direction of jet motion is a must to show up as a misalignment in the jet direction,   
projected in the sky plane, as seen by the observer. 

Figure 1 shows the geometry of the bend in the jet. Originally the jet is along OA, lying in the plane ZOX, 
making an angle $\theta$ to the observer's line of sight, assumed to be along OZ. 
The sky plane is defined by YOX. The jet undergoes a bend at point A and is moving thereafter along AB making an angle $\zeta$ to the original direction 
OAC. The plane ABC is defined by the azimuth angle $\phi$ with respect to the plane ZOC (which is the same as the plane ZOX). Our goal here is to 
determine  $\theta_1$, the angle between AP and AB, as it is $\theta_1$ that would determine the relativistic beaming factor of the jet after the bending.

\begin{figure}[t] 
\includegraphics[width=\columnwidth]{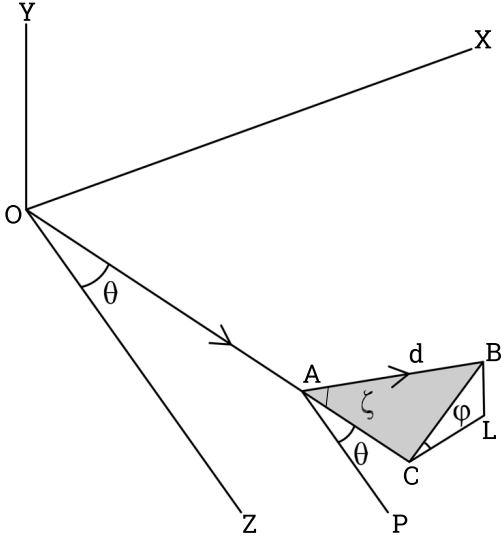}
\caption{The geometry of the bending in the jet}
\end{figure}
To the observer, the original direction of the jet in the sky plane YOX will appear to be along OX. The misalignment $\zeta$ will be the angle that 
the projection of vector AB on the sky-plane makes with OX. Vector AB is broken into a component of length $d \cos\zeta$ along AC and 
a perpendicular component along BC of length $d \sin\zeta$, the latter in turn giving components $d \sin\zeta\cos\phi$ along CL and  
$d\sin\zeta\sin\phi$ along BL. The component of AB along OX therefore is 
$d(\cos\zeta\sin\theta+\sin\zeta\cos\phi\cos\theta)$ while that along OY is $d\sin\zeta\sin\phi$. 
Therefore the misalignment in the jet direction, as seen by the observer (with line of sight along PA) is given by,
\begin{equation}
\label{eq:p59.1}
\tan \eta = \frac{\sin\zeta\sin\phi} {\cos\zeta\sin\theta+\sin\zeta\cos\phi\cos\theta}.
\end{equation}
This expression for jet misalignment is the same as derived by Conway \&  Murphy (1993).

Of course what decides the jet brightness is the orientation angle $\theta_1$ between the 
observer's line of sight and the intrinsic direction of jet motion after the misalignment.  
From Fig.~1, we need to determine projection of AB along AP. The two components along AC and CL give projection along AP as  
$d \cos\zeta\cos\theta$ and $-d \sin\zeta\cos \phi\sin\theta$ respectively. Thus angle $\theta_1$ as a function of  $\zeta, \phi$ 
and $\theta$ is given by the expression,
\begin{equation}
\label{eq:p59.2}
\cos\theta_1=\cos\zeta\cos\theta-\sin\zeta\cos\phi\sin\theta.
\end{equation}
From Eq.~(\ref{eq:p59.1}) we can express the intrinsic bending angle $\zeta$ in terms of $\eta, \theta$ and $\phi$ as,
\begin{equation}
\label{eq:p59.5}
\tan \zeta = \frac{\tan\eta\sin\theta} {\sin\phi-\tan\eta\cos\phi\cos\theta}.
\end{equation}
Then using Eq.~(\ref{eq:p59.2}), one can compute the corresponding $\theta_1$ for this bending angle. 
\section{BRIGHTNESS CHANGES WITH ORIENTATION ANGLE}
A relativistic jet with a velocity $v=\beta c$ (and a corresponding Lorentz factor $\gamma=1/\sqrt{1-\beta ^2}$), moving 
along an orientation angle $\theta$ (with respect to the line of sight in the observer's frame),  
has a beaming $\delta ^{n+\alpha}$ with Doppler factor 
$\delta = 1/(\gamma (1-\beta \cos \theta))$ and $\alpha$ the spectral index defined as $I_\nu \propto \nu^{-\alpha}$. 
Beaming becomes large as $\theta$ becomes small; $\delta = \gamma$ when $\sin \theta = 1/\gamma$. 
As for the index $n$, one should use $n=2$ if one is considering 
the integrated jet emission. This is because due to the time compression for the approaching component, a life time $\tau$ in the intrinsic frame
will have a shorter duration $\tau/\delta$ in observer's frame. Thus with a lesser number of components visible at any time,  
the integrated emission will also be less. However if one is considering the jet brightness (i.e. flux density per unit solid angle), 
then one should use $n=3$ in the beaming formula, as we shall be doing here.

It is to be noted that the beaming factor becomes unity when 
$\sin \theta = \sqrt{2/(1+\gamma)}$, and in fact for still larger $\theta$ it becomes less than unity, with $\delta = 1/\gamma$ for 
$\theta=\pi /2$. Therefore for say, $\gamma=10$, the brightness will be reduced for observers seeing the jet at right angles by a factor 
$10^{n+\alpha}$. Thus relativistic jets lying in the sky plane, the observed jet brightness may be many orders of magnitude 
weaker than its intrinsic brightness in the rest frame. 

If a jet is observed as heavily beamed then, being close to line of sight ($\sin \theta \approx 1/\gamma$), we do not normally expect it 
to show large changes in the orientation angle $\theta$ as that would change the beaming by a large factor, causing a large drop in the jet 
brightness. Therefore, if anything, large changes in $\theta$ should appear more like gaps in the jet. Here we are neither going into the physics of 
jet formation nor entertaining the question what might cause such large bends in a highly relativistic flow (see Appl et al. 1996); 
we are only examining expected changes in 
its apparent brightness if such large misalignments do take place. For brightness comparison it does not matter whether the bend is sharp or 
gradual, what matter are the initial and final orientation angle values ($\theta$ versus $\theta_1$).

If bending makes the orientation of the jet to a different value $\theta_1$, then the Doppler beaming factor would change to $\delta_1^{3+\alpha}$ 
where $\delta_1$ is the Doppler factor corresponding to $\theta_1$, i.e., $\delta_1 = 1/(\gamma (1-\beta \cos \theta_1))$. That means the observed 
brightness of the jet will change by a factor $(\delta_1/\delta)^{3+\alpha}$.  Actually the observed brightness 
of the jet would change by another factor, $\sin\theta/ \sin\theta_1$, which is a pure geometric projection effect. This projection factor is 
not accounted for in the relativistic beaming formula and is 
independent of the motion of the jet. The assumption here is that the jet is an optically thin linear feature, and when observed at an angle $\theta$, 
due to geometric projection the length perceived will be foreshortened by a factor $\sin \theta$, therefore its apparent brightness 
will be higher by a factor $1/\sin \theta$. The ratio of the jet brightness after the misalignment to that before is then given by,
\begin{equation}
\label{eq:p59.4a}
B = \left[\frac{1-\beta \cos\theta}{1-\beta \cos\theta_1}\right]^{3+\alpha} \frac{\sin\theta}{\sin\theta_1}.
\end{equation}
Now the brightness ratio is unity ($B=1$), if $\theta_1=\theta$, which from Eq.~(\ref{eq:p59.2}) will happen when
\begin{equation}
\label{eq:p59.2a}
\tan (\zeta/2) = -\tan\theta \cos\phi.
\end{equation}
 
\section{RESULTS AND DISCUSSION}
Pearson \& Readhead (1988) noted that the distribution of misalignment angles in a core-dominated sample is bimodal with one peak near $0^\circ$ 
(aligned sources) while another peak around $90^\circ$ (misaligned or orthogonal sources).  Conway \& Murphy (1993) as well as Appl et al. (1996) 
also found the distribution of misalignment angles to be bimodal with the secondary peak again around $90^\circ$. More recently Kharb et al. (2010) 
in an independent sample found the distribution to be a smooth one with only a marginal peak around $90^\circ$. In any case misalignments of 
$60^\circ$ or larger are found in $\sim 45-50\%$ of all these cases, and these large misalignments are often seen without any large changes in jet 
brightness. 

Could such misalignments appear large purely as a result of projection effects? The prevalent notion in the literature is that even though we 
see large misalignments in the jets, actual bendings ($\zeta$) may be small and because of observer's line of sight being at small angle to the jet 
(a prerequisite for large relativistic beaming), even small intrinsic bending may appear as large misalignment due to the geometry of projection 
(cf. Eq.~(\ref{eq:p59.0})). 
It is thought that since actual bending of the jets are very small, any changes in the relativistic beaming effects may also be small and large 
brightness changes do not occur. We shall show the fallacy of this notion. For one thing, arguments leading to Eq.~(\ref{eq:p59.0}) are true only for 
a specific case of $\phi=90^\circ$, but in reality $\phi$ has equal probability of being any value between 0 and $180^\circ$. Even otherwise,  
what really decides the beamed intensity is the orientation angle $\theta_1$ that the misaligned part of the jet makes with the line of sight 
of the observer, where even a small change from the erstwhile orientation angle $\theta$ could make huge difference in the jet brightness. 
\begin{figure}[t] 
\includegraphics[width=\columnwidth]{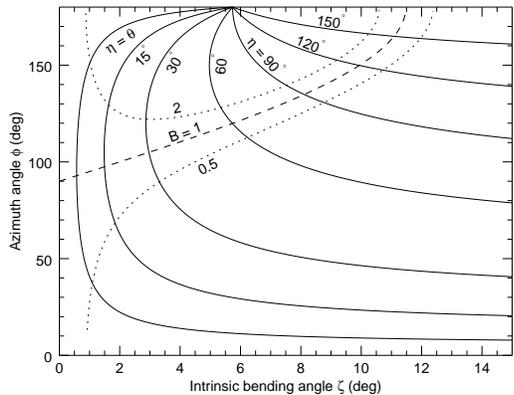}
\caption{Tracks of various misalignment angles ($\eta$) in the $\zeta, \phi$ plane. The dashed curve represents no change in the brightness ($B=1$)  
after a misalignment, while dotted curves mark the boundary where brightness changes by a factor of two.  The initial orientation angle $\theta$ of 
the jet with respect to observer's line of sight is assumed to be $\sin^{-1} (1/\gamma)$, with $\gamma=10$.}  
\end{figure}
\begin{figure}[t] 
\includegraphics[width=\columnwidth]{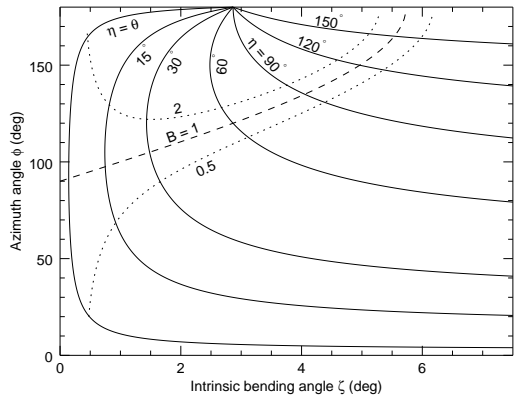}
\caption{The same as Fig. 2 but with $\gamma=20$.}  
\end{figure}

The problem is actually two-folds. Firstly, it is difficult to get a population that will give a peak in the misalignment angle $\eta$ at 
around $\sim 90^\circ$. Conway and Murphy (1993) showed that if $\phi$ is randomly distributed (as it should be because it is the angle 
between two completely independent planes) for no distribution of $\zeta, \theta, \gamma$ one could get a peak in the misalignment 
angle $\eta \sim 90^\circ$. Thus observed misalignments are difficult to obtain. 
Secondly, even if we ignore the difficulty of getting the observed distribution from any viable statistical distribution,  
and concentrate on individual 
cases of large misalignment angles (which after all can be obtained for some specific chosen values of $\zeta, \phi, \theta$ etc.), then we 
may still have to match the observed relative brightness of the jets before and after the bending with the values expected from relativistic boosting, 
and this might be an equally daunting task. 
\begin{figure}[t] 
\includegraphics[width=\columnwidth]{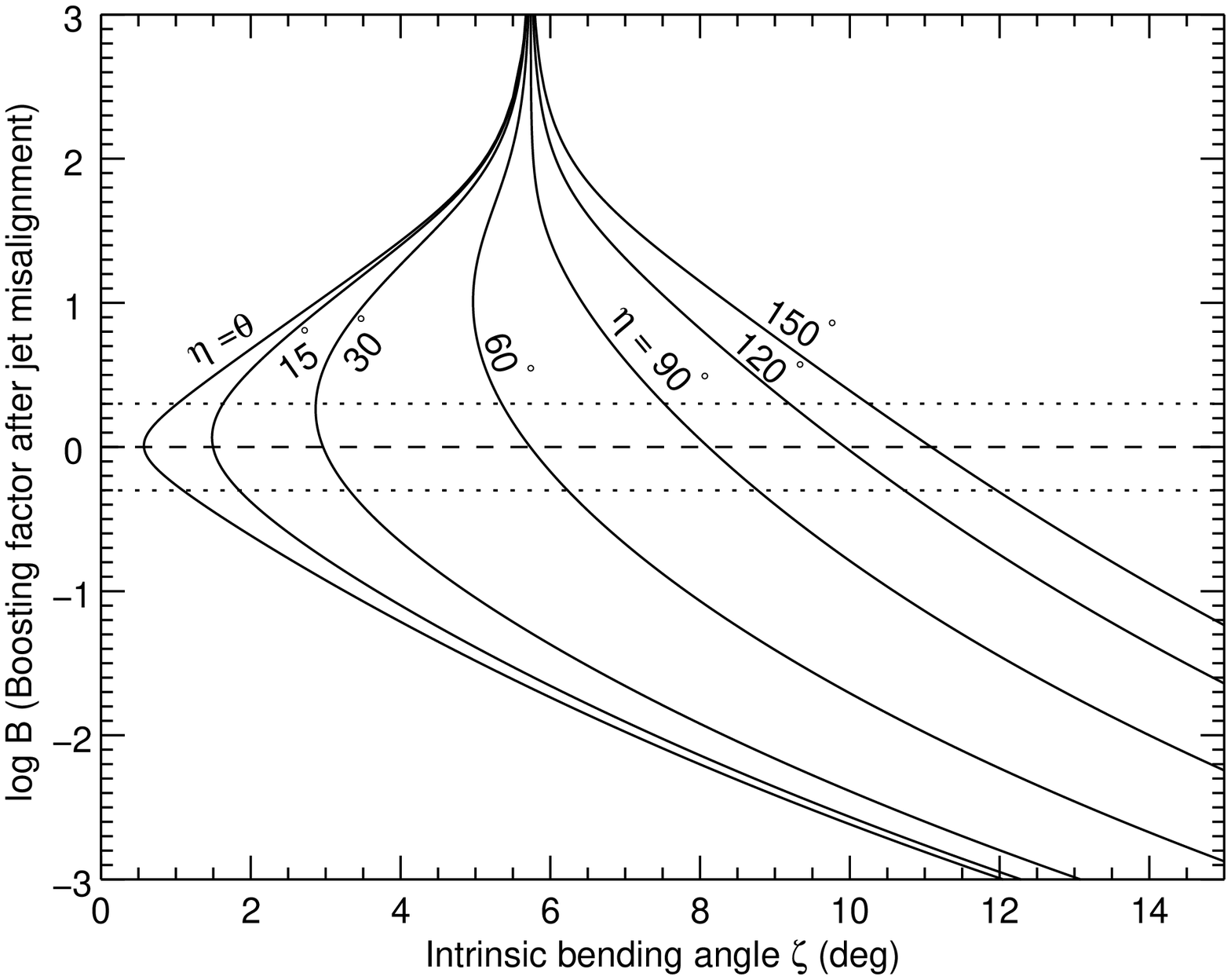}
\caption{The change expected in the jet brightness after different misalignment values ($\eta$), 
as a function of the intrinsic jet bending angle $\zeta$. The initial orientation angle $\theta$ of the jet with 
respect to observer's line of sight is assumed to be $\sin^{-1} (1/\gamma)$, with $\gamma=10$. 
The dashed line represents no change in the brightness while dotted lines 
mark the boundary where brightness changes by a factor of two.}  
\end{figure}

The bending geometry is particularly simple for a misalignment $\eta=90^\circ$, where the distribution shows a second peak. From Eq.~(\ref{eq:p59.5}) 
we can write, $\tan \zeta = -{\tan\theta}/{\cos\phi}$,
the negative sign implying $\phi> 90^\circ$ since $\theta$ and $\zeta$ are presumably small. 
Then from Eq.~(\ref{eq:p59.2}) we get a simple relation, $\cos\theta_1={\cos\zeta}/{\cos\theta}$. If $\zeta_c$ denotes  the critical value of 
the bending angle where $B=1$ (or $\theta_1 = \theta$), then for $\eta=90^\circ$, we get $\zeta_c\approx\sqrt 2\: \theta$ (for a small $\theta$). 
Thus for $\gamma=10$, $\theta \approx 5\!\!\stackrel{^\circ}{_{\cdot}}\!\!7 $ and  we have $\zeta_c \approx 8^\circ$ for a $90^\circ$ misalignment.
Even for a small change in bending, e.g., $\zeta\approx 6^\circ$, it can be easily calculated from Eq.~(\ref{eq:p59.4a}) that the jet would 
brighten by more than two orders of magnitude ($B \approx 10^{2.5}$). A smaller $\zeta$ does not necessarily imply no change in the jet brightness.

In a more general case, assuming the jet has an initial orientation $\theta$ with respect to the observer's line of sight, any particular 
misalignment seen by the observer is determined by a set of ($\zeta,\phi$) pairs in the $\zeta-\phi$ plane.  
Figure~2 shows such a diagram for an orientation angle $\theta=\sin^{-1}(1/\gamma)$ with $\gamma = 10$. Different solid curves 
plotted are for different misalignment angles ($\eta$). The dashed curve represents no change in brightness after a misalignment 
(i.e., $B=1$ or $\theta_1=\theta$), and its intersection  with any given misalignment curve gives the critical bending angles 
($\zeta_c,\phi_c$) corresponding to no change in brightness after that misalignment. Any departures from 
the critical ($\zeta_c,\phi_c$) values, would imply $\theta_1 \ne \theta$ and could result in large brightness changes (Eq.~(\ref{eq:p59.4a})) 
after the misalignment. Also plotted in Fig.~2 are two dotted curves showing a change in brightness by a factor of two 
after a misalignment. The idea in the literature that the intrinsic bending angle $\zeta$ may be small does not necessarily imply that there might 
be no appreciable change in the ensuing jet brightness. We also repeated this exercise with  $\theta=\sin^{-1}(1/\gamma)$ but with $\gamma = 20$ 
(Fig.~3). We find that it resulted in a scaled up version of Fig.~2, with $\zeta$ expanded by a factor of 2 (ratio of $\theta\approx 1/\gamma$ with 
$\gamma$ changing from 10 to 20), and except for a minor change in the 
$\eta=\theta$ curve, there were almost no perceptible changes for any of the plotted curves in Fig.~3 from what is seen in Fig.~2.
To get an idea of the possible changes in brightness, we have plotted in Fig.~4 the brightness change $B$ against $\zeta$ for different 
misalignments, all again for our chosen case of $\sin\theta=1/\gamma$ with $\gamma = 10$. Except for in a very narrow range around  
$\zeta_c$ (i.e., between dotted lines in Fig.~4), we get large changes in the brightness. The whole scenario of large misalignments with no accompanying 
change in brightness is very unlikely to happen.

It is not possible to calculate the exact probabilities as we have no idea about the values $\zeta$ in a jet could take. However, from observed 
$\eta$ one can get some constraints on the possible values of $\zeta$. From Figs. 2 and 3 it can be seen that
for $\zeta \ll \theta$, there will hardly be any change in the brightness, due actually to a very small change in the orientation angle 
$\theta_1 \approx \theta$. 
But then the misalignment angle seen in the jet also cannot be large, i.e., $\eta  \stackrel{<}{_{\sim}} \theta$. However with large misalignments 
($\eta \sim 90^\circ$ or higher) often seen in jets, the bending angle $\zeta \ge \theta$, as also noted by Conway \& Murphy (1993). 
From Figs. 2 and 3 we see that the brightness ratio of the jet could be much below unity ($B \ll 1$) for large bending angles 
($\zeta > 2 \theta$), and which could very well happen as $\zeta$ and $\theta$ are quantities completely independent of each other. 
As for the azimuth angle $\phi$, we can 
be sure that $\phi$ is a random variable between $0$ and $\pi$ as it is an angle between two independent planes, one determined by 
the intrinsic bending of the jet and the other determined by the observer's line of sight. A small range of $\phi$ between the two dotted lines in 
Figs. 2 and 3 means that only a small percent of the cases one expects to see brightness changes within a factor of two, and that in rest  
of the cases brightness changes after the misalignments would be much larger, even if we assume $\zeta$ range to be the 
most favorable, i.e., $\zeta$ does not go beyond $2\theta$.
\begin{table}[t]
\begin{center}
\caption{Bending angles ($\zeta$) for various misalignment angles ($\eta$), and the corresponding fraction of the azimuth angle (${\Delta\phi}/{2\pi}$) 
for $0.5\le B\le2$}
\hskip4pc\vbox{\columnwidth=33pc
\begin{tabular}{ccc}
\tableline\tableline 
$\eta$ & $\zeta$ & ${\Delta\phi}/{2\pi}$ \\
(1)&(2)&(3)\\
\hline
$\stackrel{<}{_{\sim}}\theta$ & $\stackrel{<}{_{\sim}}\theta/6$ & $\stackrel{>}{_{\sim}}0.7$ \\
$15^\circ$ & $\stackrel{>}{_{\sim}}\theta/4$ & $\sim 0.3$  \\
$30^\circ$ & $\stackrel{>}{_{\sim}}\theta/2$ & $\sim 0.15$ \\
$60^\circ$ & $\stackrel{>}{_{\sim}}5\theta/6$ & $\sim 0.1$  \\
$\ge 90^\circ$ & $>\theta$ & $\sim 0.03-0.06$ \\
\tableline
\end{tabular}
}
\end{center}
\end{table}
\begin{table}[t]
\begin{center}
\caption{Bending angles and the relative numbers of high contrast vs. low contrast sources after misalignments (${N(|B|>2)}/{N(|B|<2)}$)}
\hskip4pc\vbox{\columnwidth=33pc
\begin{tabular}{ccc}
\tableline\tableline 
$\zeta$ &  ${\Delta\phi}/{2\pi}$ & ${N(|B|>2)}/{N(|B|<2)}$ \\
(1)&(2)&(3)\\
\hline
$\zeta\ll\theta$   & $\stackrel{>}{_{\sim}}0.8$ & $<1$\\
$0.2\theta\stackrel{<}{_{\sim}}\zeta<\theta$  & $\sim 0.2-0.5$ & $\sim1-5$ \\
$\theta\stackrel{<}{_{\sim}}\zeta\stackrel{<}{_{\sim}}2\theta$ & $\sim 0.03-0.1$ & $\sim 10-30$ \\
$\zeta>2\theta$ &  $< 0.01$ & $>10^2$\\
\tableline
\end{tabular}
}
\end{center}
\end{table}

\begin{figure}[t] 
\includegraphics[width=\columnwidth]{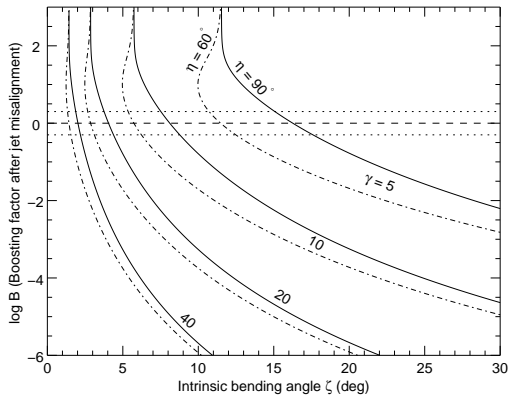}
\caption{The change expected in the jet brightness after an observed misalignment  
as a function of the intrinsic jet bending angle $\zeta$, for various $\gamma$ values. The initial orientation angle $\theta$ of the jet with 
respect to observer's line of sight is assumed to be $\sin^{-1} (1/\gamma)$. The family of curves plotted are for two different 
misalignment angles, $\eta=90^\circ$ -- solid curves, $\eta=60^\circ$ -- dot-dash curves. The horizontal dashed line represents no change in the 
brightness while the horizontal dotted lines mark the boundary where brightness changes by a factor of two.}  
\end{figure}
\begin{figure}[t] 
\includegraphics[width=\columnwidth]{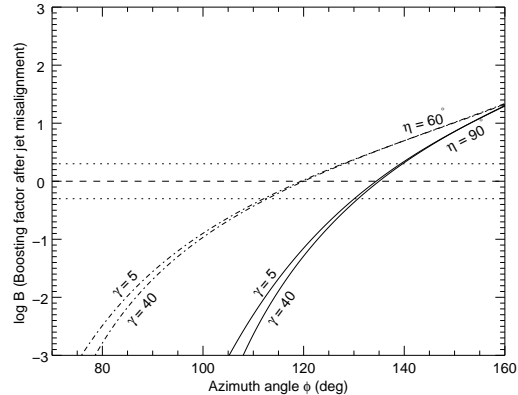}
\caption{The same as Fig. 5, but now the change expected in the jet brightness after an observed misalignment plotted 
as a function of the azimuth angle $\phi$, for various $\gamma$ values.}  
\end{figure}
The results are summarized in Table 1, which is organized in the following manner: 
(1) Misalignment angle ($\eta$).
(2) Bending angle ($\zeta$).
(3) The fraction of the azimuth angle (${\Delta\phi}/{2\pi}$) for $0.5\le B\le2$. 
We may point out that most entries are approximate numbers, to indicate trends. Although we have no inkling of the distribution of possible 
values $\zeta$ might take, yet it is still possible to get some idea of (${N(|B|>2)}/{N(|B|<2)}$) for different ranges of $\theta$, irrespective of the 
misalignments $\eta$. Table 2 shows that, which is organized in the following manner: 
(1) Bending angle ($\zeta$).
(2) The fraction of the azimuth angle (${\Delta\phi}/{2\pi}$) for $0.5\le B\le2$. 
(3) Number of sources with  brightness contrast larger than two, as compared to those with  contrast smaller than two (${N(|B|>2)}/{N(|B|<2)}$). 
Of course this also has to be kept in mind, that $\zeta$ 
and $\theta$ are otherwise completely independent 
quantities, while $\zeta$ is something intrinsic to the jet and its value may get determined by the jet physics or local circumstances near the 
location of the bend, $\theta$ is a pure chance value decided by line of sight of the observer and the jet axis.  

Figure 5 shows a plot of brightness change as a function of $\zeta$ for the misalignment angle $\eta=90^\circ$ and $\eta=60^\circ$, 
$\theta=\sin^{-1} (1/\gamma)$ for various $\gamma$ values. What one again sees is that a small change in the bending angle (from $\zeta_c$ 
to $\zeta \approx \theta$), can make the jet after the bending brighter by many orders of magnitude. 
Figure 6 shows a plot of beaming factor against azimuth angle $\phi$ for the misalignment angle $\eta=90^\circ$ and $\eta=60^\circ$, 
for $\theta=\sin^{-1} (1/\gamma)$ for various $\gamma$ values. We see that while critical value of $\phi$ does depend upon the misalignment 
angle, it is more or less independent of the Lorentz factor $\gamma$ of the jet motion and we see that our overall conclusions do not change for 
different but still large misalignment angles, i.e. for $\eta \gg \theta$.

Of course, we assumed no change in the intrinsic brightness 
and we also assumed no change in the speed of the jet material, only change assumed is in the direction of the jet. 
This was done to keep the problem simple and the number of free parameters to be a minimum.
Even otherwise, to assume just right amount of changes in the intrinsic properties of the jet or in its relativistic speed 
so as to cancel neatly any variation in the relativistic beaming factor due to the change in the orientation angle, 
for it to appear as a result with the same 
brightness after the bending as it was before, would be a rather contrived scenario. 

The question of observed jet brightness comparison on either side of misalignments has not been systematically explored in the literature. 
A quantitative comparison of the flux ratios on either side of the bend 
may, however, need to be corrected for numerous selection effects. As we discussed above, there would be many more large 
misalignments with large flux ratios than could be missed because of difficulties in measuring flux ratios of jets differing in brightness by more than 
an order of magnitude because of dynamic range limitations. In a proper, carefully selected sample of bending angles, observed with sufficiently 
good sesitivity, the distributions of the brightness 
ratios on either side of the bends would need to be consistent with the predictions made here, if relativistic beaming is true. Presently such data are  
either not yet available or not in a form to directly test or resolve the issues raised here. We may point out that there are independent arguments in 
the literature (Bell 2012) that Doppler boosting may have played no significant role in the finding surveys of radio-loud quasars.
\section{CONCLUSIONS}
We have shown that the relativistic beaming models along with the observed large misalignments seen in the jets of active galactic nuclei,  
predict large contrasts in the brightness observed before and after the misalignments. It was also shown that  
for every large misalignment ($\zeta \stackrel{>}{_{\sim}} 60^\circ$) detected, there might be an order of magnitude larger number of similar 
misalignments which might not have been seen because of high brightness ratios. That would also imply that large misalignments occur an order of 
magnitude or more than what have been inferred observationally. Carefully selected samples of jet misalignments, with measured brightness ratios 
of the jet brightness on either side of the bends, would be needed to test the consistency of the relativistic beaming hypothesis observationally.

\end{document}